\documentclass[usenatbib]{mn2e}
\usepackage{times}
\usepackage{epsfig}
\newcommand{\eg}{e.$.\!$g.\ }
\newcommand{\be}{\begin{equation}}
\newcommand{\ee}{\end{equation}}
\newcommand{\etal}{et al.\ }
\newcommand{\chandra}{{\it Chandra} }
\newcommand{\xmm}{{\it XMM-Newton} }

\newcommand{\ie}{i$.\!$e$.\!$, }

\newcommand{\ki}{$\chi^2$\,}

\title[X-ray and optical study of a fossil cluster]
{A fossil galaxy cluster; an X-ray and optical study of RX J1416.4+2315}
\author[Khosroshahi \etal]{
Habib G. Khosroshahi$^1$\thanks{E-mail:
habib@star.sr.bham.ac.uk}
Ben J. Maughan$^2$\thanks{Chandra Fellow}, Trevor J. Ponman$^{1}$ \&  
Laurence R. Jones$^{1}$ \\
$^{1}$School of Physics and Astronomy, The University of Birmingham,
Birmingham B15 2TT, UK\\
$^{2}$Harvard-Smithsonian Center for Astrophysics, 60 Garden St, Cambridge, 
MA 02138, USA.}

\begin{document}

\date{Accepted, Received}

\pagerange{\pageref{firstpage}--\pageref{lastpage}} \pubyear{2005}

\maketitle

\label{firstpage}

\begin{abstract} 
We present a detailed X-ray and optical study of a distant fossil
system RX J1416.4+2315 ($z\approx 0.137$), combining \chandra and \xmm
observations, optical photometry and spectroscopy. X-ray emitting hot
gas imaged by both the \chandra and \xmm shows a globally relaxed
spatial distribution, supporting the idea that fossil groups are old
galaxy systems with no recent mergers. However, the diffuse X-ray
emission shows signs of asymmetries in the core of the system. With a
mean gas temperature of $\sim 4.0$ keV and total gravitational mass of
3.1$\times 10^{14} M_{\odot}$, within the virial radius, this is
better described as a fossil galaxy cluster rather than a fossil
group. The temperature profile shows no sign of a significant cooler
core despite a cooling time dropping to 5 Gyr within the resolved
core. We find a mass concentration parameter $c_{200}\sim 11$ which is
relatively high for a cluster of this mass, indicative of an early
formation epoch.

Using the spectroscopically identified cluster members we present the
galaxy luminosity function for this fossil system. We measure the
velocity dispersion of the galaxies to be $\sim 700$ km s$^{-1}$ based
on 18 confirmed members. The dynamical mass is nearly twice the total
gravitational mass derived from the X-ray analysis. The measured
R-band mass-to-light ratio, within the virial radius, is $\sim 440$
M$_\odot$/L$_\odot$ which is not unusual for clusters of galaxies. The
central giant elliptical galaxy has discy isophotes and spectral
features typical of elliptical galaxies.

\end{abstract}


\section{Introduction}

Though rich clusters of galaxies are the host for giant elliptical
galaxies, their high velocity dispersion is less conducive to
galaxy-galaxy interactions and mergers. In contrast, smaller galaxy
systems with low velocity dispersions are more merger efficient, and
hence potential sites for the formation of bright and massive
galaxies. N-body simulations of clusters in a hierarchical
cosmological model show that galaxy merging naturally produces a
massive, central galaxy with surface brightness and velocity
dispersion profiles similar to observed brightest cluster
galaxies. The central galaxy forms through the merger of several
massive galaxies along a filament early in the cluster's history.

If a merger efficient system such as a galaxy group remains
undisturbed due to relative isolation from other massive systems, a
giant elliptical galaxy will form as a result of the internal multiple
merger due to dynamical friction. The time-scale of such a process
depends on the local density and is shorter at early formation
epochs. The resulting system should lack $L^\star$ galaxies but the fainter
end of the luminosity function should remain intact, as only a small
fraction of the accreted mass is due to the dynamical friction of
smaller galaxies over a Hubble time \citep{dubinski98}.  With highly
relaxed intra-group hot gas revealed by X-ray observations and a gap
of at least 2 magnitude between the first and second ranked galaxies
\citep{jones03}, galaxy groups known as ``fossil groups'' appear to be 
the observed products of the above mechanism \citep{ponman94}.

There are about a dozen such systems identified \citep{ponman94,
vikh99, mz99, jones00, romer00, mat01, jones03} but few of them have
been studied in detail \citep{kjp04,sun04}. Based on the space density
of fossils, \citet{jones03} estimate that fossil systems represent
8\%-20\% of all galaxy systems with the same X-ray luminosity, and are
as numerous as poor and rich clusters combined. Therefore they account
for a significant fraction of the galaxy systems in the universe,
though observationally they are difficult to find, especially at
higher redshifts.

About half a dozen fossils have been studied in the X-ray where they
are most impressive.  The nearest known fossil group NGC 6482, for
instance, has shown interesting properties \citep{kjp04}, such as an
unusually high mass concentration and absence of a cooler core despite 
the very short cooling time. \citet{sun04} studied a more massive fossil
using \chandra and \xmm, and find a much smaller cool region than
expected. If our arguments about the early formation epoch of the
fossils are correct, these systems should be ideal laboratories to
study cool cores. This is because of the absence of an obvious cooling
flow removal mechanism, major cluster merging, and therefore other
heating mechanisms can be looked into with less ambiguity.

Fossils groups are equally important in their optical properties, as
they provide a suitable environment to study the stellar properties of
giant elliptical galaxies. Fossils are isolated and old systems
enabling us to make a direct comparison between the morphology of
ellipticals grown in fossil groups and their counterparts at the core
of galaxy clusters. This is further motivated by an earlier finding
that there are enough fossils, from their space density, to provide
the giant dominant ellipticals in clusters. There is a need for
detailed optical studies to address various issues such as the
mass-to-light ratio and luminosity function of galaxies, in addition
to X-ray studies. The central galaxy warrants special
attention, as the age of its stellar population can provide further clues 
to the cluster history, complementing the X-ray studies.
 
This paper is organised as following: Section 2 briefly reviews the
group properties and describes the observations, data reduction and
preparations. The results from imaging and spectral X-ray analysis are
presented in section 3. Section 4 describes the distribution of
mass. Results from optical and near-IR photometry, optical
spectroscopy and mass-to-light ratio are presented in section 5. A
discussion and concluding remarks are summarised in section 6.

We adopt a cosmology with $H_0=70$ km~s$^{-1}$~Mpc$^{-1}$ and
$\Omega_m=0.3$ with cosmological constant $\Omega_\Lambda=0.7$
throughout this paper. At the redshift of RX J1416+2315 the luminosity and
angular diameter distances are 650 Mpc and 500 Mpc, respectively and 1
arcsec $\equiv$ 2.44 kpc.

\section{Observation and Preparation}

\subsection{The group:}

The group RX J1416.4+2315 (hereafter J1416), $z\approx 0.137$, was
first studied as part of a volume limited sample of spatially extended
X-ray sources compiled during the WARPS project (Wide Angle ROSAT
Pointed Survey; \citep{scharf97, jones98, perlman02}). Details of the
initial sample selection and the search leading to the volume limited
sample of fossil groups can be found in \citet{jones03}. It is the
most X-ray luminous source in the sample of 5 fossil groups studied by
\citet{jones03}, with a ROSAT estimated X-ray bolometric luminosity of
$1.1 \times 10^{44}$ erg s$^{-1}$. In a deep optical image it would be
classified as a galaxy group or poor cluster centred on an extremely
dominant, luminous giant elliptical galaxy.  The very extended X-ray
emission is detected to a semi-major axis length of 3.5 arcmin (512
$h^{-1}_{70}$ kpc) in the ROSAT PSPC (Position Sensitive Proportional
Counter) and is elongated in a direction similar to that of the
central elliptical galaxy.

This galaxy cluster appears in the bright SHARC cluster survey of
\citet{romer00}. The redshift is confirmed, but it is not identified as 
a fossil system by \citet{romer00}. The target of this ROSAT field was
a candidate cluster of galaxies, but at a high redshift (z$>$ 0.3),
and unrelated to the fossil system.

The VLA FIRST catalog of radio sources gives a radio source at the
centre of the giant elliptical galaxy with an integrated flux of 3.39
mJy at 1.4 GHz. The radio source is extended on a scale of 4 arcsec.

\begin{figure}
\begin{center}
\scalebox{0.3}{\includegraphics*[angle=270]{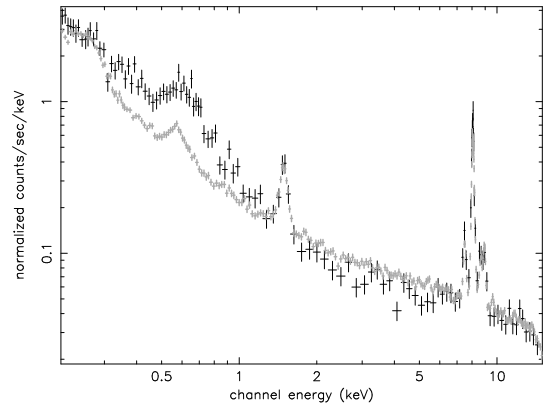}} \\
\caption{\label{fig.bgspec}Background spectra obtained from the source
field (black) and blank sky dataset (grey). The blank sky spectrum was
normalised by the flux of events detected outside the field of view.}
\end{center}
\end{figure}

\subsection{\chandra observations} 

The \chandra observation of this group was performed in September 2001
(Obs ID 2024) for a total of 14.8 ks using \chandra ACIS-S. No major flares 
were detected therefore an exposure of 14.5 is used in the analysis. Though
this observation helps to detect and position the point sources with a
high spatial accuracy and gives enough counts to study the global
properties such as the spatial distribution and the core temperature
of the hot gas, it does not provide enough counts for a more detailed
analysis out to a large radius.

We follow the same procedure as \citet{kjp04} for the
\chandra analysis except that we use a later version of the analysis 
tools (CIAO 3.2 and CALDB 3.0). We therefore correct for the quantum
efficiency degradation of ACIS, which increases with time. We use CTI
corrected blank-sky background files in our imaging and spectral
analysis as the ACIS-S3 chip is contaminated by the source emission
due to the large extension of the X-ray halo. To study the soft
diffuse X-ray emission, point sources were detected using CIAO
``wavdetect'' and removed. Their locations were then filled with
surrounding background using a linear interpolation.  A total of 20
point sources were detected. The brightest point source was an
off-centre background AGN at 14:16:24.7, +23:16:14.9 (J2000).

\subsection{\xmm observations}\label{sect.xmmprep}

The large collecting area and field of view of \xmm make the \xmm
observations attractive despite relatively poor spatial resolution
compared to that of \chandra.  The system was observed by \xmm for
28ks in July 2003. The observation was split into two parts of 19ks
and 9ks.  The data were reduced and analysed with version 6.1 of the
Science Analysis System (SAS) and with the most recent calibration
database as of January 2005. The 19ks observation was rendered
completely unusable by extremely high background levels, though the
second interval was mostly unaffected by high background
periods. After cleaning, there were $8$ks and $4$ks of good time for
the MOS and PN cameras respectively. Blank sky background datasets
\citep{read03} were also prepared and cleaned in the same way.

The reasonably small size of the source on the EPIC detectors meant
that there was sufficient local background data to enable useful
comparisons to be made with the blank sky background fields to
determine their suitability for data analysis purposes for this
dataset. The spectrum of the background in the observation was compared to that
in the blank sky data, after normalising them by the particle background
flux measured from the events detected outside the field of view of the
telescopes. The resulting background spectra are plotted in Fig 
\ref{fig.bgspec} for the PN detector (the MOS spectra are similar).

There is a significant excess in the source dataset at the soft end, due
to higher Galactic X-ray background at the source position. The
normalisation of the continuum above 2 keV is too high in the blank sky
background, but the normalisation in the particle-induced fluorescent
lines (1.5 and ~8 keV) is correct. This means that the ratio of particle
to hard X-ray background flux in the source data set is higher than that
in the blank sky background. When the blank sky spectrum is normalised by
the particle background flux, the other background components are
overestimated, giving rise to a higher continuum level.

\begin{figure}
\center
\epsfig{file=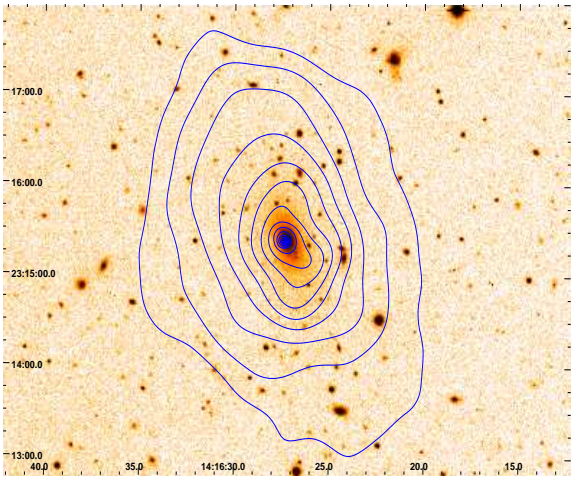,width=3.0in,height=2.5in}
\caption{X-ray contours from the soft (0.3-2.0 keV) diffuse, point sources 
removed, X-ray emission detected by \chandra are overlaid on a 
6$\times$6 arcmin$^2$ size deep R-band optical image. }
\label{chansoft}
\end{figure}

This result suggests an alternative method of normalising the blank sky
background; using the $2-7$ keV flux. The upper limit is imposed to avoid
another fluorescent line complex in the PN data. This method gives the 
correct $>2$ keV continuum, but underestimates the fluorescent line flux. 
The effect of these two methods on temperature measurements is discussed 
later.

\subsection{Optical photometry and spectroscopy} 
 
The imaging and spectroscopic observations of the group was performed
using the observational facilities of the Issac-Newton Group of
Telescopes (ING), Kitt-peak National Observatory (KPNO) and UK
Infrared Telescope (UKIRT). The R-band image was obtained using the
INT 2.5m wide field imager during a service time observation in August
2000. Unfortunately the conditions were not photometric, so further
R-band imaging was obtained with the 8k mosaic camera at the
University of Hawaii 2.2-m telescope in photometric conditions, and
used to calibrate the original images.

The spectroscopic observations were performed during a run to study
several fossil groups using a multi-slit spectrograph on the KPNO-4m
telescope on the 11-13th March 2000. The Ritchey-Cretien spectrograph
and KPC-10A grating gave a dispersion of 2.75 $\AA$/pixel over
3800-8500 $\AA$ and with 1.8 arcsec slitlets a resolution of 6$\AA$
(FWHM) was achieved. Risley prisms compensated for atmospheric
dispersion. Spectra were obtained through three slitmasks, with
typically an hour exposure on each. The spectroscopic data were
reduced and analysed in the standard way using IRAF.

The high quality near infrared images of the target were obtained in
June 2004 using UIST/UKIRT. Data reduction was performed using the
Observatory Reduction and Acquisition Control (ORAC) data reduction
tool and four 180 sec exposure mosaic images were then combined to increase
the signal-to-noise.

\begin{figure}
\center
\epsfig{file=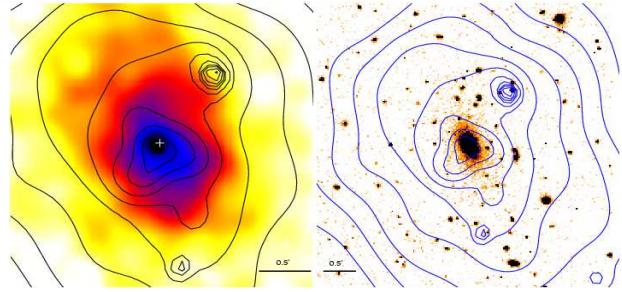,width=3.2in,height=1.5in}
\caption{X-ray soft (0.5-2 keV) diffuse emission contours from \xmm 
overlaid on \chandra soft diffuse X-ray emission (left) and the 
R-band image of the central galaxy (right). 
Both the \chandra and \xmm show asymmetries in the core. The white cross 
shows the location of the radio source from the FIRST catalog and matches  
the position of a point source detected at the centre of the cluster.}
\label{xmmcont}
\end{figure}

\section{X-ray analysis and results}

\subsection{Spatial distribution}

Figure \ref{chansoft} shows the contours of soft (0.3-2.0 keV) diffuse
X-ray emission from the \chandra observation on R-band optical
image. The contours extended to 500 kpc along the semi-major axis of
the X-ray emission. They appear relaxed and aligned with the stellar
major axis of the giant elliptical galaxy.
 
The \xmm soft (0.5-2 keV) diffuse X-ray emission contours are also
shown (Fig. \ref{xmmcont}). The diffuse emission image was produced by
adaptively smoothing an exposure-corrected image of the emission
detected by the three \xmm EPIC cameras to $99\%$ significance. Unlike
Fig. \ref{chansoft}, here the point sources are not excluded. The
off-centre AGN can be seen just North-West of the centre of the X-ray
emission. The X-ray emission from the system at this energy range
appears relaxed, overall, except in the core, where asymmetries are
seen. The \chandra diffuse emission of the core shows a NE-SW
elongation and a SW 'tail' (Fig. \ref{chansoft}) which are also seen
in the \xmm contours (Fig \ref{xmmcont}). However, some of the other
irregularities seen in the \xmm image are not apparent in the
\chandra data. 

\begin{figure}
\center
\epsfig{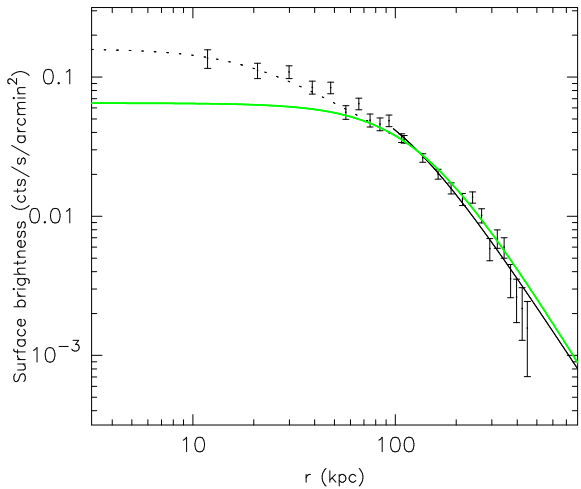}
\caption{Radial surface brightness profile and the double $\beta$-model fit 
to the soft energy (0.3-2.0 keV) X-ray emission from \chandra. The dotted 
curve is the $\beta$-model fit to the inner region, $\sim$100 kpc. The
dark solid curve is the $\beta$-model to the outer region. The thick 
gray (green) curve represents the $\beta$-model fit to the \xmm surface 
brightness profile.}

\label{rprof}
\end{figure}

\subsubsection{Surface brightness distribution}

A radial surface brightness profile (Fig \ref{rprof}) was extracted 
from the inner 200 arcsec of the ACIS-S3 image in a soft energy range 
(0.3-2.0 keV) and fitted by a one-dimensional $\beta$-model,

\be
\Sigma(r)=\Sigma_0[1+(\frac{r}{r_0})^2]^{-3\beta+1/2},
\ee
where $r_0$ and $\beta$ are the core radius and index, and
$\Sigma_0$ the central surface brightness. 
This gives a core radius of $r_0$= $28.2_{-4.43}^{+5.14}$ arcsec 
and $\beta$=$0.47_{-0.02}^{+0.03}$ but is far from a good fit to the data. 

With the clear ellipticity in the two-dimensional X-ray distribution
and larger area covered by \xmm EPIC we performed a two dimensional
$\beta$-model fit to the MOS 1, MOS 2 and PN images at the same energy
band. The core radius along the major axis is $r_0=54\pm4$ arcsec, with
$\beta=0.54\pm0.02$ and ellipticity of
$0.34\pm0.06$. A two-dimensional $\beta$-model fit to the \chandra data gives 
a $r_0 = 36\pm3$ arcsec and $\beta = 0.44\pm 0.01$ with an ellipticity 
of 0.36.  
 
There is a large difference in the values of $\beta$ and the core
radius obtained from \chandra and \xmm analysis. To find the root of
the discrepancy and also because of a relatively poor fit to the
\chandra data we performed a double-$\beta$ fit by fitting inner and
outer regions of the image each with a single $\beta$-model. The 
parameters of these $\beta$-models are, however, tied to give a
continuous gas density \citep{kjp04,pratt02} at the break radius, 100 kpc. We
find $r_{0in}$= $7.7_{-3.1}^{+3.3}$ arcsec and
$\beta_{in}$=$0.31_{-0.03}^{+0.06}$ and $r_{0out}$=
$37.0_{-4.3}^{+4.1}$ arcsec and $\beta_{out}$=$0.53_{-0.02}^{+0.02}$
with a reduced $\chi^2$=1.32. Fig \ref{rprof} shows this $\beta$-model
fit and the $\beta$-model fit to the \xmm surface brightness profile.

\begin{figure}
\center \epsfig{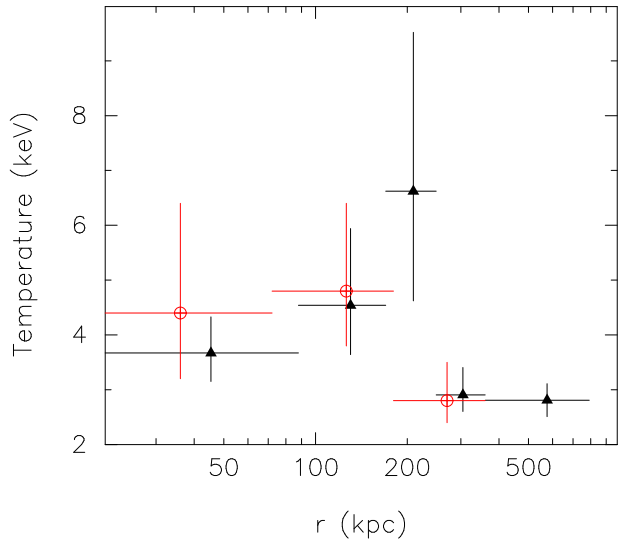}
\caption{Temperature profile of J1416. The triangles (black) show  
the deprojected temperature profile from \xmm analysis 
and the circles (red) is the same from \chandra analysis.}
\label{tprofile}
\end{figure}

\subsection{Spatially resolved spectroscopy}

\subsubsection{\chandra}

The ACIS spectra in three successive circular annuli were extracted
and fitted with a de-projected (using XSPEC projct model)
absorbed APEC \citep{smith01} model. A fixed hydrogen column
density of $N_{H, gal}=2\times 10^{20}\, {\rm cm}^{-2}$ was
included in the model to account for Galactic absorption. The
background was extracted for each annulus from the blank sky
observations in the same detector regions as the source spectra. Each
spectrum contains about 1000 counts in bins of 20 counts each. Since
we were not able to estimate the metallicity from the spectral fits, we
fixed the metallicity to 0.5 $Z_\odot$, the mean value from \xmm
analysis on a similar scale.

The unabsorbed flux from the central 200 arcsec is $1.22\times10^{-12}$
erg s$^{-1}$ cm$^{-2}$, which corresponds to an X-ray luminosity,
1.04$\times10^{44}$ ergs/s (0.5-8.0 keV) and a bolometric luminosity of 
1.24$\times10^{44}$ ergs/s.  The fitted column 
density was found to be consistent with the galactic level, though 
poorly constrained with an error of about $0.1\times10^{22}$ cm$^{-2}$.

\subsubsection{\xmm}

Spectra were extracted from 5 circular annular regions defined to
contain $\ge 1200$ source counts. The effect of different background
subtraction methods on the measured temperatures was investigated. A
local background spectrum was extracted from an annular region free
from source emission. A background spectrum was extracted from the
blank sky background in the same detector region as the source
spectrum, and was corrected for the residuals between the local and
blank sky backgrounds using the ``double subtraction'' method
described by (\eg \citet{arnaud02}). These two methods were also
repeated with the events weighted (using evigweight) to correct for
the telescope's vignetting. This weighting has the disadvantage of
incorrectly boosting the non-vignetted particle background
component. The two blank sky methods (\ie with and without weighting)
were performed with the blank sky background normalised by either the
flux of events detected outside the field of view of the telescopes or
using the 2-7 keV flux. This gives a total of 6 background subtraction
methods.

The spectra from each detector were grouped to contain $\ge 20$ counts per
energy bin, and were fit simultaneously in the 0.4-7keV band by an
absorbed APEC model. The absorbing column was fixed at the Galactic value
as before, and the temperature, metal abundance and normalisation were
free parameters in the fit. The temperature and metallicity of the PN and
MOS spectral models were tied together, but the normalisations were
allowed to fit independently.

The temperature profiles obtained with the non-weighted methods were
found to be in excellent agreement, regardless of background
subtraction method. However, the methods using weighted spectra were
less consistent.  As one might expect, the dispersion of temperatures
increased with radius, as the background becomes more important. The
agreement between the non-weighted methods is excellent at all radii,
while the temperatures measured with the (evig) weighted methods
become inconsistent with each other, and the non-weighted methods at
larger radii. This is perhaps unsurprising, as we have already seen
that the non-vignetted particle background in the source dataset is
relatively high (section \ref{sect.xmmprep}). Based on these results,
we chose to use the non-weighted, local background method for all of
the \xmm spectral analysis.

The surface brightness distribution of J1416 is clearly elliptical, so
the effect of this on the measured quantities such as the temperature
profile is investigated by extracting spectra in elliptical
annuli. The ellipses were chosen to closely match the isophotes, and
contain $\sim1200$ source counts. The resulting temperature profile
was consistent with that measured with circular annuli, and so the
circular temperature profile is used throughout for consistency with
our assumption of spherical symmetry in estimating the mass of the
system.

With the methodology thus established, the temperature profile was
deprojected using the XSPEC projct model. Similarly we obtained a
PSF-deconvolved temperature profile, however due to limited statistics
we were unable to constrain the temperatures with the choice of bin
sizes described above which is the largest to enable us perform a mass
profile analysis. We therefore did not use the PSF-deconvolved
temperature profile for this study.

\subsection{Temperature and metallicity}
Fig \ref{tprofile} shows the deprojected temperature profile of
the cluster based on various analysis methods out to 4 arcmin ($\sim
700$ kpc). The \xmm temperature agrees with the overall temperature
found from \chandra observations of the central 130 kpc. Within the
resolution of the data we find no significant drop in the temperature
of the central annulus, in contrast to the observed temperature
profiles of many relaxed clusters of galaxies which present cool cores
\citep{vikh05}.

With limited counts from \chandra observations, our estimation of the
metal abundances are based on the \xmm observations, however, the
deprojected metallicities are very poorly constrained. The mean
metallicity, within 400 kpc, is found to be 0.23$\pm$0.076 solar metallicity.

\subsection{Gas entropy and cooling time}

\begin{figure}
\center
\epsfig{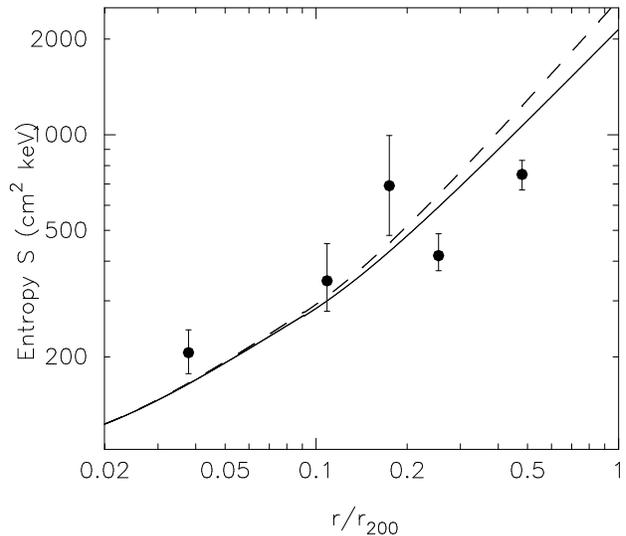}
\caption{Entropy profile of the hot gas in J1416. The data points 
are obtained from emission measure for each annulus with its associated 
temperature. Solid and dashed
lines are the entropy profiles assuming polytropic and isothermal
temperature profiles, respectively. }

\label{scent}
\end{figure}

The entropy of the X-ray emitting gas is defined here as:

\be
S(r)=kT(r)/n_e(r)^{2/3}\, {\rm keV cm}^2 ,
\ee
where $n_e$ is the electron density. The shape of the entropy profile
is of special interest since preheating models predict large
isentropic cores. Using the $\beta$-model fit to the surface
brightness profile and the central gas density, obtained from the
spectral fit, we can derive an analytical relation for the gas entropy
providing that we model the temperature profile using an analytical
function. Two entropy profiles shown in Fig \ref{scent} are obtained
in this way by assuming isothermality ($T\approx 4.0$ keV) and a
polytropic fit ($\gamma=1.1$) to the temperature profile. In addition
we obtain the gas entropy in each spherical shell (data points in Fig
\ref{scent}) by estimating the electron density via the emission measure
assuming a Raymond-Smith plasma emission model for each spherical
shell and the $L$ is the X-ray luminosity from the shell with the
volume $V$. The gas entropy at $0.1r_{200}$ (see section 3.5.1 for
estimate of $r_{200}$) is $S_{0.1}\approx300\pm50$ keV cm$^2$ similar
to the average value obtained by \citet{psf03} for clusters with similar temperatures. No isentropic core 
is seen, though our spatial resolution is limited. This is consistent with 
other examples of group and cluster scale systems reported to
show no large isentropic cores \citep{pratt03,psf03}.

\begin{figure}
\center
\epsfig{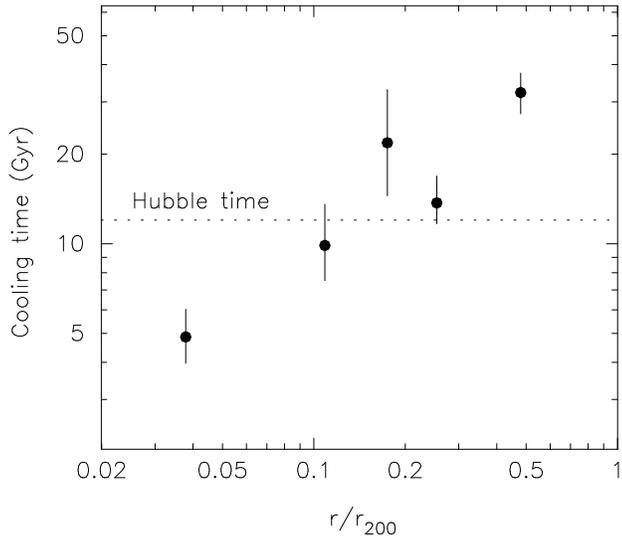}
\caption{The estimated cooling time for each shell. The resolved core has 
a cooling time $\sim 5$ Gyr, yet we observe no cooler core in the 
temperature profile.}
\label{tcool}
\end{figure}

The cooling time of the gas was calculated by dividing the total
thermal energy of each shell by the associated emissivity, 
$t_{cool}\sim\frac{3}{2}\frac{(n_e+n_H)kT}{\epsilon}$, where $\epsilon$ is 
the bolometric emissivity obtained by assuming a Raymond-Smith plasma 
emission model. The resulting profile is presented in Fig \ref{tcool}. 

\subsection{Gas and total gravitational mass profiles}

The gas mass and the total gravitational mass profile can be derived
from the gas density and the temperature profiles assuming that the
gas is in hydrostatic equilibrium and is distributed with spherical
symmetry. To support our assumptions we note that the distribution of
the hot gas in this system is fairly symmetric and regular with
the exception of a small region in the core. In addition, the
comparison between the temperatures from circular and elliptical
shells shows no significant difference in this case. The \chandra analysis 
of the X-ray surface brightness distribution shows that there is no 
significant discrepancy in the results of the surface brightness profile 
by performing a one-dimensional analysis by selecting circular annuli or 
a two-dimensional fit. Finally \citet{fabricant84} show that the mass 
profile, from X-ray analysis, of an oblate or prolate system is similar 
to the one obtained under spherical symmetry assumption.

The total gravitational mass is given by:
\be
M_{grav}(<r)=-\frac{kT(r)r}{G\mu m_p}\left[\frac
{dln\rho(r)}{dlnr}+\frac{dlnT(r)}{dlnr}\right].
\ee
where G and $m_p$ are the gravitational constant and proton mass and
$\mu=0.6$.

The temperature profile seen in Fig \ref{tprofile} is modelled by a
polytropic model with $\beta$ and core radius fixed to the values
obtained from the surface brightness fit to the \chandra data giving
$\gamma\simeq 1.09$. It is then
straightforward to derive an analytical description for the total mass
using the expressions for density under the polytropic
assumption. However the error in the mass should
be obtained indirectly.  We use a Monte-Carlo method which is described in
\citet{kjp04} and \citet{nb95} for our error estimation.  In brief,
the formula for the mass profile consists of two parts, the variation in gas
density ($d\ln n/d\ln r$) and in the temperature ($T(r)$ and $d\ln T/d\ln
r$). For the latter we have generated 1000 physical temperature
profiles with a temperature 
envelope defined by the observed errors in the 
temperature and linear interpolation. A physical temperature
profile is one which guarantees a monotonically increasing mass with radius.
In this simulation the density profile parameters were fixed at the
values from the $\beta$-model fit.  In order to estimate the
contribution of the density variation ($d\ln n/d\ln r$) to the mass
error, the surface brightness profile at the location of each data
point was fitted with $d\ln n/d\ln r$ as the free parameter instead of
$\beta$. The error was estimated at the 68\% confidence level using
the error matrix provided by the fitting program. We checked the
results by monitoring the \ki variation. The total error in the mass at
each data point was then derived by adding the two uncertainties
quadratically.

\begin{figure}
\center
\epsfig{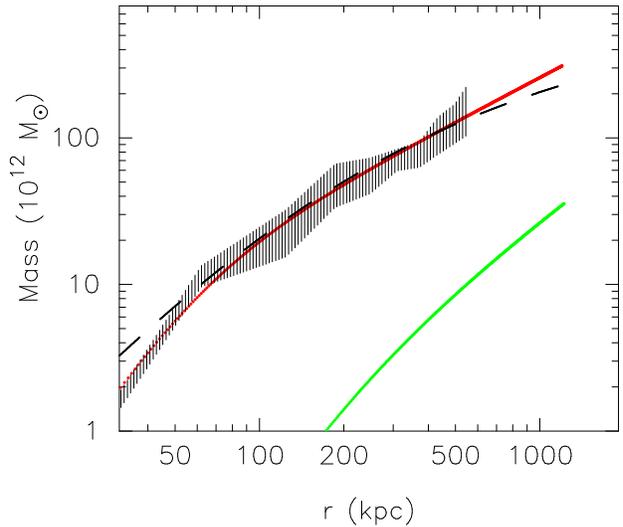}
\caption{Total gravitational mass and gas mass profiles of J1416. 
The upper solid (red) curve shows the gravitational mass profile.
The dashed curve represents the NFW profile.   
The error envelope of the total gravitational mass is shown
with vertical hashes. The lower solid (green) curve shows the 
gas mass profile.  The concentration parameter is $c_{200}=11.2\pm4.5$. 
The mass data points and errors are those estimated using MC 
simulations. }
\label{massprof}
\end{figure}

\subsubsection{NFW mass profile and concentration}

Motivated by the results of numerous cosmological N-body simulations
and to enable us to make a direct comparison with mass profiles from
other studies, we attempt to fit a NFW profile \citep{nfw95}

\be
\rho_m(r)=\frac{4\rho_m(r_s)}{(r/r_s)(1+r/r_s)^2},
\ee
to the total gravitational mass density.
Integrated the mass profile for a spherical mass distribution leads to, 
\be \label{nfw}
M_{tot}(<r)=16\pi\rho_m(r_s)r_s^2[r_s \ln (1+r/r_s)-\frac{r}{1+r/r_s}],
\ee
where $\rho_m(r_s)$ is the density at $r_s$. The mass concentration
parameter can then be defined as $c_{200}=r_{200}/r_s$. By definition,
$r_{200}$ is the radius within which the mean gravitational mass
density is 200 times the critical density, $\rho_c(z)$. For our
assumed cosmology ($\Omega_m$=0.3, $\Lambda$=0.7), the critical
density is $\rho_c(z)=\frac{3H_0^2}{8\pi G}(1+z)^2(1+z\Omega_m)$.

We fit the NFW mass profile (equation \ref{nfw}) to the five observed
data points with errors obtained from MC simulation. The best fit
profile has $r_s=97$ kpc. $r_{200}$ is then calculated by
extrapolating the profile to $200\rho_c$. This gives $r_{200}=
1.22\pm0.06$ Mpc. The extrapolated total mass, using the best fit NFW
profile, at $r_{200}$ is $M_{200}= 3.1(\pm1.0)\times 10^{14}$
M$_{\odot}$.

To obtain the concentration of the dark matter we take into account
the contribution of the stellar mass of the central dominant galaxy,
for which we have the R-band luminosity and assume a stellar
M/L$_R\sim$ 5 M$_\odot$/L$_\odot$ from the recent fundamental plane
studies \citep{capp05}. This gives $r_s=108$ kpc and therefore a
concentration parameter $c_{200}= 11.2\pm4.5$, which is high compared
to the values of $c_{200}$ for poor clusters studied by
\citet{pratt05}. Implications of this high mass concentration is discussed 
in section 5.

The integrated gas mass can be derived by integrating the
gas density. Fig \ref{massprof} shows the gas mass profile. 
The total gas mass within $r_{200}$ is $\sim 4\times 10^{13}$ M$\odot$. With 
the extrapolated total gravitational mass within the same radius 
$\sim 3.1\times 10^{14}$ M$\odot$, the gas fraction is calculated to be 
$\sim 0.13$. 

\section{Optical photometry and spectroscopy}

To estimate the completeness of our spectroscopic observation we extract 
galaxies down to $R=19.5$ mag using Sextractor. This gives a total of 48
sources within 0.5$r_{200}$ (the radius used in the definition
of fossil systems by \citet{jones03}), of which 9 are
classified as stars. From the total of 39 galaxies, only 27 were 
observed, of which 18 were cluster members with a mean
redshift of 0.137 and calculated galaxy velocity dispersion of 
$\sigma=700\pm120$ km s$^{-1}$. 
Some of the galaxies (8 in total) have been missed due to the slit 
configuration. Four galaxies were know to be foreground galaxies prior 
to this observation. We estimate the completeness of the sample to be 77 
percent within 0.5$r_{200}$ and down to $m_R=19.5$. Based on 
this calibration, the magnitude of the central galaxy is 15.17 and the 
second brightest galaxy in the cluster is $m_R=17.7$. None of the galaxies
missed due to the slit configuration were brighter than $R=17.7$, thus the 
system does meet the formal fossil criteria of \citet{jones03}.

\subsection{Luminosity function}
\begin{figure}
\center \epsfig{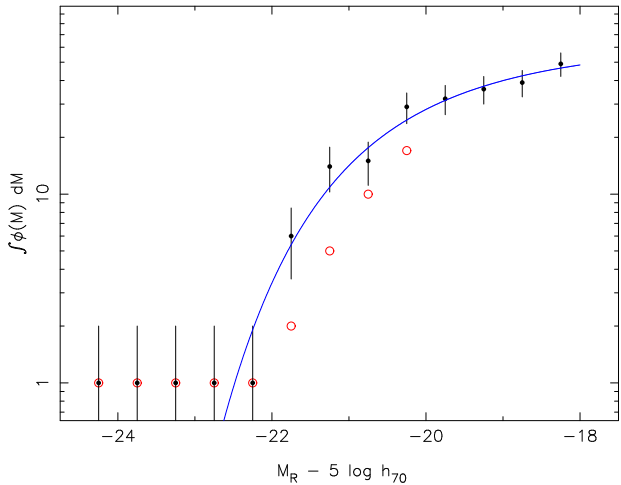}
\caption{Galaxy luminosity function of J1416. The dark data points and the 
associated error bars show the luminosity function of the galaxies using 
a statistical subtraction of the background using two regions outside 
the virial radius. The curve (blue) is the best fit Schecter function to 
these data points. The open circles (red) represent the  luminosity function 
of the the confirmed cluster members.}
\label{glf}
\end{figure}

The luminosity function (LF) of galaxies in this fossil cluster is
obtained by selecting the galaxies down to $M_R\sim$-18.5. The
cumulative LF is shown in Fig \ref{glf} after statistical subtraction
of the background galaxies extracted from two fields around the source
just outside the $r_{200}$ of the cluster. In this figure we show the
luminosity function of the spectroscopically confirmed cluster members 
(open red circles) and the luminosity function of galaxies (filled circles) 
after a statistical subtraction of the field galaxies and the associated 
error bars.

Fitting a Schecter function to the LF of confirmed members, we
obtain  $\alpha= -1.23\pm0.28$ and $M^\star_R= -20.40\pm0.22$. 
An independent study of a higher quality galaxy sample for
this fossil cluster results in a similar slope \citep{mendes06}. 
We obtain $\alpha= -0.61\pm0.20$ and $M^\star_R=
-20.55 \pm0.40$ when we fit a Schecter function to the LF of
photometrically selected galaxies (solid curve in the Fig \ref{glf}). 
As this figure shows, this luminosity function is considerably
shallower. There is an inconsistency in the slopes of the galaxy 
luminosity functions of the confirmed members and that
obtained from  statistical background subtraction, which could be
due to the fact that the statistical background subtraction is 
performed based on the galaxy classification, using SExtractor, in the 
absence of galaxy colours. 

For RX J1552.2+2013, another fossil with a similar mass to J1416, 
\cite{mendes05} find a slope of $\alpha=-0.77\pm0.37$ for the 
spectroscopically selected galaxies and $\alpha=-0.64\pm0.30$ for the
photometrically selected galaxies. 
From the study of the GEMS X-ray groups, \citet{miles04} find $\alpha=-1.0$.
They also report a dip in the luminosity function which is more prominent in 
poor groups and is argued to be the result of galaxy mergers in
groups. A similar mechanism is thought to be the origin of the large
gap in fossils' luminosity function. This is discussed in section 5.

\subsection{Mass to light ratio }

Having the total gravitational mass and the optical luminosity of the
cluster we are in a position to estimate the mass-to-light ratio of
the system.

The total R-band luminosity of the cluster is measured to be
10$^{11.85}$ $L_{\odot}$ by accumulating the luminosity of the
photometrically identified cluster members within $r_{200}$. Thus the
total mass-to-light ratio within $r_{200}$ is $\sim 440$
$M_{\odot}/L_{\odot}$. Converting this to the mass-to-light ratio in
B-band we find $M/L_B\sim 580$ $M_{\odot}/L_{\odot}$. This $M/L$,
derived assuming B-R=1.5, is an upper limit because all the cluster
galaxies are assumed to be early-types. Under a similar assumption
\citet{mendes05} find a mass-to-light ratio of $M/L_B\sim 750$
$M_{\odot}/L_{\odot}$ for another massive fossil based on a dynamical
mass estimate. These mass to light ratios, ours and that of
\citet{mendes05}, are relatively high compared to the results of 
distant clusters $M/L_R \sim 210 M_\odot/L_\odot$
\citep{carlberg96,girardi02}. However, higher values of mass to light 
ratio are also reported \citep{mohr96}. In comparison to other fossils
\citep{vikh99,yoshioka04} these fossils and NGC 6482 \citep{kjp04} have 
lower mass-to-light ratio and hence difficult to argue whether or not
fossils have, in general, higher values of $M/L$.

\subsection{The central galaxy}

\begin{figure}
\center \epsfig{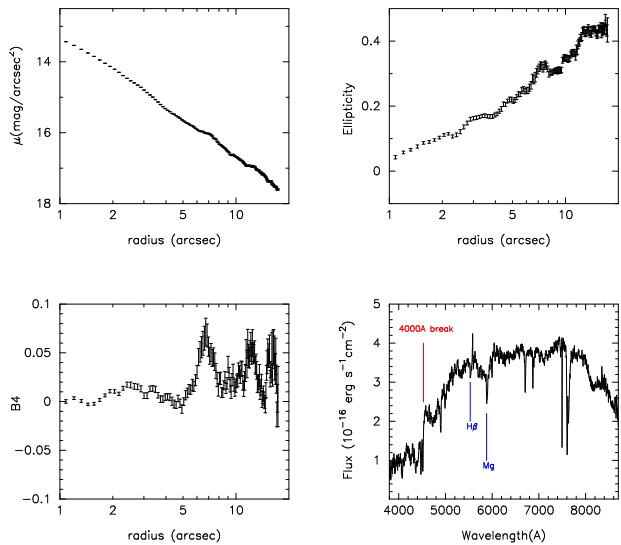}
\caption{Isophotal analysis of the central galaxy in K-band. Surface 
brightness radial profile (top-left), ellipticity profile (top-right), 
Cosine coefficient B4 (bottom-left) and the stellar spectrum of the 
giant elliptical galaxy are presented. }
\label{isophot}
\end{figure}

To study the stellar light distribution of the central giant
elliptical galaxy we performed an isophotal analysis using the high
resolution UIST K-band images of the system. The isophotal analysis
using the IRAF task 'ellipse' shows a discy nature for the stellar
light at an outer radius of $r> 20$ kpc as shown in figure Fig
\ref{isophot}. The ellipticity of the galaxy increases to a value of
0.43, overtaking the ellipticity of the X-ray distribution as
expected \citep{buote96}.  
To find out if this is a pure elliptical galaxy we fitted a
Sersic profile ($r^{1/n}$) to the radial surface brightness profile
using a two dimensional decomposition technique which allows for a
possible disc component, with an exponential profile. We find no disc
contribution to the total galaxy light. We find a
Sersic index of $n\sim3.0$ and a half light radius of $\sim 20$
kpc. The giant elliptical galaxies in galaxy clusters often show boxy
isophotes \citep{bender92} while having larger values of $n$
\citep{habib04}. Further discussion should be based on a well 
defined sample of dominant galaxies in fossils, which is the subject
of a separate study.

The spectrum of the central galaxy is similar to that of a typical
elliptical galaxy, with $H\beta$ and $Mg$ lines marked in Fig \ref{isophot}.

\section{Discussion and Conclusions}

Using the high angular resolution of the \chandra and higher
sensitivity of the \xmm, combined with optical imaging and
spectroscopy and near IR imaging, we studied the most massive and hot
fossil galaxy system, which also hosts the the most luminous giant
elliptical galaxy observed in fossils.

We confirm the result of an earlier ROSAT study that RX J1416.4+2315
meets the requirements of being a ''fossil'' discussed in
\citet{jones03}. These are the minimum X-ray luminosity of $10^{42}$
erg s$^{-1}$ and a minimum of two magnitude difference in R-band
luminosity between the two brightest galaxy belonging to the system (within 
0.5$r_{vir}$). The examination was necessary as the gas temperature 
in the earlier study was underestimated by $\sim 40\%$ resulting in 
a smaller virial radius within half of which we had to examine the above 
optical criteria. 

The global X-ray distribution is relaxed, except a small region at the
core, indicating an absence of any recent group-group merger and major
galaxy-galaxy merger indicative of an early formation epoch. The
central dominant galaxy is a giant elliptical galaxy aligned with the
X-ray emission and hence the dark matter distribution. The alignment
which is also seen in NGC 6482 \citep{kjp04} and ESO 3060170
\citep{sun04} is an interesting feature which is known to be the
consequence of an anisotropic collapse, such as along a filament. The
alignment effect only works for the central giant elliptical galaxy
and is seen in the N-body simulations
\citep{dubinski98}. Recent hydrodynamical cosmological simulations, to 
investigate the effect of baryonic dissipation on halo shapes
\citep{kzd05}, show that there is also a strong correlation between dark
matter halo shape and stellar remnant morphology in collisionless
mergers.

Despite the high degree of relaxation and symmetry, there are signs of
asymmetry at the core of the system. These are seen in the X-ray
soft emission within the core. This could be
a result of AGN activity. VLA FIRST catalog gives the integrated
flux of the central AGN in the giant elliptical galaxy as 3.39 mJy
at 1.4 GHz, with a spatial size of 4 arcsec. Radio observations of
the system are currently being performed to investigate the detailed
morphology of the radio emission and the status of the AGN duty cycle.

If fossil groups are old systems, as argued, they should provide ideal
environments for the formation of cool cores, due to the absence of
any recent major merger which can remove the cool core. The
temperature profile shows no significantly cooler core, suggesting the
existence of a heating mechanism. Given the limited statistics and the
\xmm point spread function, the presence of a small ($<50$ kpc) cooler region 
cannot be ruled out. However the absence of cool core is also found in
the case of the nearest observed fossil, NGC 6482 \citep{kjp04} where
the core is resolved with 1 kpc resolution. In addition ESO 3060170
\citep{sun04} shows a cool core much smaller than the expected cooling
region for a system of few $10^{14}$ $M_\odot$. As mentioned above
there is evidence of AGN activity in the core of RX J1416+2315, similarly
to ESO 3060170.

This study shows that the X-ray temperature from a previous ROSAT
study was underestimated by about half its value. It is therefore
important to study the effects of these changes on the scaling
relations involving X-ray quantities. This is the subject of a
separate study concentrating on the X-ray scaling relations of 
fossil galaxy groups.

The initial argument for the early formation of fossils relied on the
relaxed X-ray morphology, absence of recent mergers and the large gap
(2 mag) in the optical luminosity function of their constituent
galaxies. Their mass distribution, and in particular their mass
concentration parameter, can provide more solid and independent
evidence for their early formation. \citet{wechs02} show that halos
which have not had a major merger since z = 2 are more concentrated in
comparison to systems which had experienced a major merger since
then. We measure $c_{200}$ and find that the mass concentration for
this system is relatively high compared to clusters of similar mass
\citep{pratt05}. This is excluding the stellar mass contribution to the 
total gravitational mass from the central giant elliptical galaxy. We
have already reported a very high mass concentration parameter for NGC
6482 \citep{kjp04}. In addition, \citet{sun04} find a high mass
concentration ($\sim 8.9$) when they fit a NFW profile for ESO
3060170, another massive fossil. If fossils are old systems, their
core must have formed in an early stages when the density of matter
was higher, resulting in formation of a dense and compact core whilst
later accretion of mass builds an extended halo. Relatively higher
mass concentration of low mass systems compared to more massive halos
are seen in cosmological simulations \citep{bullock01}.

The central galaxy in this cluster has discy isophotes despite the
fact that majority of giant elliptical galaxies at the cores of
clusters and groups show boxy isophotes \citep{kb96}. There are other
examples of discy isophotes in fossils' central galaxies. NGC 6482
also shows highly discy isophotes. The original fossil, RX J1340+4017
has pure elliptical isophotes. We are currently studying the
possibility that this, non-boxy isophotes, is a generic property of
giant ellipticals at the core of fossil systems.

A more accurate galaxy luminosity function, compared to the first
published for a fossil (Jones etal 2001), is presented in this study.
The confirmed members form a LF which is steeper than most of the
galaxy luminosity functions presented for the groups. We find 
$\alpha= -1.23\pm0.28$ and $M^\star_R= -20.40\pm0.22$, by fitting
the Schecter function to the LF of the spectroscopically confirmed
member galaxies. This agrees well with the results of an independent 
study of this cluster by \citet{mendes06}.

\citet{miles04} found, in their study of GEMS groups, an interesting
feature referred to as ``dip'' in the optical luminosity function of 
dim (low velocity dispersion) X-ray groups compared to bright (high
velocity dispersion) X-ray groups interpreted as being produced as a
result of more efficient galaxy merger in low velocity dispersion
groups. Similarly, the main argument behind the lack of bright galaxies 
in fossils is the merger of $L^\star$ galaxies, but fossils are in general 
more X-ray luminous than the dim X-ray groups studied by \citep{miles04}. 
The velocity dispersion of galaxies in 
this fossil cluster is $\sigma\approx 700$ km s$^{-1}$ which is consistent 
with the value expected from the observed cluster $L_X$-$\sigma$ 
\citep{mahdavi01} given the X-ray luminosity of this cluster. Similarly 
J1552.2+2013, lies on the same $L_X$-$\sigma$ 
relation \citep{mendes05}, though the scatter in the $L_X$-$\sigma$ 
relation is large. 

An especially interesting feature of RX J1416.4+2315 is the large size
and mass of the system. Dynamical friction, the key process believed
to be responsible for the orbit decay which allows the major galaxies
to merge within fossil groups, is much less effective at high
velocities, raising the question of whether the existence of such a
fossil cluster poses a problem for the whole model of fossil system
formation through galaxy merging. \citet{jones00} estimated the
time-scale of dynamical friction for the original fossil group,
J1340.6+4018, and showed that $L^\star$ galaxies, initially in circular
orbits at half the virial radius and circular velocity of
$\sqrt{2}\sigma$, should merge into the central galaxy in
$4.5\times10^{9}$ yr. The subject of this study, J1416, is more
massive and has a radial velocity dispersion, $\sigma$ almost twice
that of J1340.6+4018. It also has a larger virial radius. The time
scale for orbital decay of a body of mass $M$ by dynamical friction is
proportional to $\frac{r^2 v_c}{M}$, \citep{binney87}, which means
that a $L^\star$ galaxy in this system will require a dynamical friction
time-scale larger than a Hubble time to merge. We note, however, that
this estimate is based on initial conditions in which the galaxy is
part of a virialised system and rotates on a circular orbit around the
centre. If instead, galaxies fall into the developing cluster along a
filament, the loss of the angular momentum could occur in a much
shorter time, due to the small pericentre radius of their orbits. The
X-ray emission, and hence the dark matter distribution in J1416, is
highly elongated, implying a very anisotropic velocity dispersion
tensor, and supporting the idea of collapse along a dominant filament.
 
Recent simulations of the formation of fossil systems also suggest
that they can be formed with high masses.
\citet{donghia05} find that fossil groups have already assembled half of 
their final mass at z$\sim$1, and subsequently they typically grow by
minor merging only. The early assembly of fossils groups leaves
sufficient time for $L^\star$ galaxies to merge into the central galaxy by
dynamical friction, resulting in the large magnitude gap.

We would like to thank \chandra and \xmm observatory teams and 
the anonymous referee for helpful comments and suggestions that improved 
the presentation of the paper. We also would like to thank John Mulchaey 
for his involvement in spectroscopic observations of the target and 
Issac-Newton group of telescopes for service observation of the target 
with INT. BJM is supported by NASA through Chandra Postdoctoral Fellowship Award Number
PF4-50034 issued by the Chandra X-ray Observatory Center, which is operated
by the Smithsonian Astrophysical Observatory for and on behalf of NASA
under contract NAS8-03060.

\bsp

\label{lastpage}
\end{document}